\documentclass{llncs}
\usepackage{graphicx}
\usepackage[usenames,dvipsnames,svgnames,table]{xcolor}
\usepackage{fancyvrb}
\usepackage{hyperref}
\usepackage{listings}

\definecolor{linkblue}{RGB}{0,0,155}

\hypersetup{
   bookmarksnumbered,
   colorlinks={true},
   pdfstartview={FitH},
   citecolor={linkblue},
   linkcolor={linkblue},
   urlcolor={linkblue},
}

\begin{document}

\title{\texttt{nanopub-java}: A Java Library for Nanopublications}

\author{
Tobias Kuhn\inst{1,2}
}

\institute{
  Department of Humanities, Social and Political Sciences, ETH Zurich, Switzerland\\
\and
  Department of Computer Science, VU University Amsterdam, Netherlands
\smallskip\\
  \texttt{kuhntobias@gmail.com}
}

\maketitle

\begin{abstract}
The concept of nanopublications was first proposed about six years ago, but it lacked openly available implementations. The library presented here is the first one that has become an official implementation of the nanopublication community. Its core features are stable, but it also contains unofficial and experimental extensions: for publishing to a decentralized server network, for defining sets of nanopublications with indexes, for informal assertions, and for digitally signing nanopublications. Most of the features of the library can also be accessed via an online validator interface.
\end{abstract}

\section{Introduction}

This technical paper presents \texttt{nanopub-java}, which is a library for nanopublications. Its source code can be found here:
\begin{quote}
\url{https://github.com/Nanopublication/nanopub-java}
\end{quote}
Nanopublications\footnote{\url{http://nanopub.org}} \cite{groth2010isu,mons2011naturegen} are an approach to publish scientific data and meta-data in RDF by subdividing them into small data snippets.
They are a concrete proposal to implement the visions of \emph{semantic publishing} \cite{shotton2009lp} and \emph{linked science} \cite{kauppinen2012dis} by allowing for the publication and sharing of formally represented scientific resources and data that are semantically interlinked and provide provenance and context information for their reliable integration and evaluation.
Specifically, a nanopublication consists of three named graphs of RDF triples (plus a fourth graph to keep them together): the assertion graph contains the actual content of the nanopublication (e.g. a scientific finding); the provenance graph contains information about the provenance of the assertion (e.g. the scientific method with which the assertion was derived); and the publication information graph contains meta-data about the nanopublication itself (e.g. its creator and a timestamp).

The library presented here can be useful in a number of scenarios:
\begin{itemize}
\item To represent and share small chunks of scientific knowledge and metadata in RDF in a provenance-aware manner (as nanopublications)
\item To make RDF content verifiable and immutable (with trusty URIs)
\item To define large or small datasets of RDF content where the data entries can be individually addressed and recombined in new datasets (with nanopublication indexes)
\item To quickly publish RDF snippets in a verifiable and permanent manner (relying on an existing server network)
\item To retrieve existing nanopublications from the network (5 millions and counting)
\item To digitally sign RDF snippets (though this is still experimental)
\end{itemize}
Below the details of the library and its web interface are explained.

\section{Implementation}

The library is built upon the Sesame library \cite{broekstra2002iswc} to validate, represent, and create RDF structures. The features of the nanopublication library are centered around a Java interface representing a nanopublication, called \texttt{Nanopub}, and the Java class \texttt{NanopubImpl} provides a reference implementation of this interface. This implementation checks the well-formedness of a nanopublication at the time of its creation based on the latest version of the nanopublication guidelines,\footnote{\url{http://nanopub.org/guidelines/working_draft/}} and raises an exception in the case of a violation of these rules.

Trusty URIs \cite{kuhn2014eswc,kuhn2015tkde} are the recommended way of how to make nanopublications verifiable and immutable, and to give them identifiers based on cryptographic hash values. The nanopublication library uses for that purpose the \texttt{trustyuri-java} library.\footnote{\url{https://github.com/trustyuri/trustyuri-java}} In a nutshell, a trusty URI is a kind of URI reference that contains a cryptographic hash value that is calculated on the digital artifact it represents. This allows one to verify that a given content is really what the URI was supposed to represent by its creator, and thereby to enforce the immutability of digital artifacts such as nanopublications.

The features of the library are made available through the Java API as well as via a command line interface using the command \texttt{np}. The following features are part of the core of the library, which means that they deal with stable and agreed-upon structures as defined by the community:

\begin{itemize}
\item \texttt{check / CheckNanopub} reads a nanopublication or several of them and checks whether any of the well-formedness criteria are violated. If a trusty URI or a digital signature is found (see below), these are checked too.
\item \texttt{mktrusty / MakeTrustyNanopub} takes a nanopublication that does not yet have a trusty URI and transforms it into one that is identified by a newly created trusty URI.
\item \texttt{fix / FixTrustyNanopub} takes a nanopublication with a broken trusty URI and fixes it, i.e. assigns it a new trusty URI. This is useful when a nanopublication has to be changed, which invalidates the hash. Running this command creates a \emph{new} nanopublication with a valid trusty URI. (Nanopublications are immutable, so changing something necessarily leads to a new nanopublication.)
\end{itemize}

In addition to these core features, the library also contains a number of unofficial extensions (which may or may not become official at some point). There is code to validate informal assertions specified as AIDA sentences \cite{kuhn2013eswc}; code for creating index nanopublications to define small or large sets of nanopublications \cite{kuhn2015iswc}; code for publishing and retrieving nanopublications from a decentralized server network \cite{kuhn2015iswc}; and experimental code for digitally signing nanopublications. These features are accessible via the following commands and classes:

\begin{itemize}
\item \texttt{mkindex / MakeIndex} takes a list of nanopublications and creates an index that refers to them. A nanopublication index therefore represents a (possibly large) set of nanopublications. Such indexes are themselves formatted as nanopublications, and can therefore also be published to the server network (see below).
\item \texttt{publish / PublishNanopub} uploads a given nanopublication that has a trusty URIs to the server network. Such a nanopublication is then distributed among the servers of the network (currently five) and made available even if some of the servers should be inaccessible at a certain point in time. In this way, the nanopublication is made permanent and its publication cannot be undone.
\item \texttt{get / GetNanopub} reliably retrieves a given nanopublication from the decentralized server network. Nanopublications are verified according to their trusty URI, and only verified nanopublications are returned by this command. For nanopublication indexes, the whole set of nanopublications that is defined by the index can be downloaded.
\item \texttt{status / NanopubStatus} checks whether and how often a given nanopublication (identified by its trusty URI) is found on the server network.
\item \texttt{server / GetServerInfo} returns some information about a given server in the network, such as the number of nanopublications it contains.
\item \texttt{mkkeys / MakeKeys} creates a new key-pair to be used to sign nanopublications.
\item \texttt{sign / SignNanopub} takes a nanopublication and signs it with a given private key.
\end{itemize}

\section{Examples}

Below, some of the most important commands are explained by examples based on the \texttt{np} command line tool. The same functionality is also available via the Java API.
For the sake of these examples, let us assume that we have a file called \texttt{nanopubfile.trig} that starts with the following RDF prefixes:
\begin{lstlisting}
{o}@prefix xsd: <http://www.w3.org/2001/XMLSchema#>.
{o}@prefix dc: <http://purl.org/dc/terms/>.
{o}@prefix pav: <http://purl.org/pav/>.
{o}@prefix prov: <http://www.w3.org/ns/prov#>.
{o}@prefix np: <http://www.nanopub.org/nschema#>.
{o}@prefix ex: <http://example.org/>.
{o}@prefix : <http://example.org/np1#>.{end}
\end{lstlisting}
The definition of the first nanopublication in this file starts with the head graph that defines the structure of the nanopublication by linking to the other graphs:
\begin{lstlisting}
{o}:Head {
{o}    : a np:Nanopublication; np:hasAssertion :assertion;
{o}        np:hasProvenance :provenance; np:hasPublicationInfo :pubinfo.
{o}}{end}
\end{lstlisting}
The actual claim of the nanopublication is stored in the assertion graph:
\begin{lstlisting}
{o}:assertion {
{o}    ex:drugA ex:treats ex:diseaseB.
{o}}{end}
\end{lstlisting}
The provenance and publication info graphs provide meta-information about the assertion and the entire nanopublication, respectively:
\begin{lstlisting}
{o}:provenance {
{o}    :assertion prov:wasDerivedFrom ex:some_publication.
{o}}
{o}:pubinfo {
{o}    : pav:createdBy <http://orcid.org/0000-0002-1267-0234>.
{o}    : dc:created "2015-08-18T15:36:22+01:00"^^xsd:dateTime.
{o}}{end}
\end{lstlisting}
The lines above constitute a very simple but complete nanopublication. To make this example a bit more interesting, let us assume that our file contains two more nanopublications that have different assertions but are otherwise identical:
\begin{lstlisting}
{o}@prefix : <http://example.org/np2#>.
{k}...
{o}:assertion {
{o}    ex:Gene1 ex:isRelatedTo ex:diseaseB.
{o}}
{k}...
{o}
{o}@prefix : <http://example.org/np3#>.
{k}...
{o}:assertion {
{o}    ex:Gene2 ex:isRelatedTo ex:diseaseB.
{o}}
{k}...{end}
\end{lstlisting}
To check and validate these three nanopublications, we can now use the following command:
\begin{lstlisting}
{c}$ np check nanopubfile.trig
{o}Summary: 3 valid (not trusty);{end}
\end{lstlisting}
These nanopublications can now be transformed into ones with trusty URIs using the following command (resulting in a new file \texttt{trusty.nanopubfile.trig}):
\begin{lstlisting}
{c}$ np mktrusty nanopubfile.trig{end}
\end{lstlisting}
Using the same command in verbose mode with the argument \texttt{-v} shows us the newly generated trusty URIs for the three nanopublications:
\begin{lstlisting}
{c}$ np mktrusty -v nanopubfile.trig
{o}Nanopub URI: http://example.org/np1#RAHGB0WzgQijR88g_rIwtPCmzYgyO4wRMT7M91ouhojsQ
{o}Nanopub URI: http://example.org/np2#RA4xTdhe2gPctqvAwdgTU4eRiR1aTQlTYJcF3Sohe5Cus
{o}Nanopub URI: http://example.org/np3#RAEjvXP0xTkeIa2mKmYT66i_PAJ-u-k0uRBd6_sMe9qG0{end}
\end{lstlisting}
As they are tiny snippets of data, nanopublications are most useful when they grouped and combined in small or large collections. We therefore need a simple method to refer to collections or sets of nanopublications, which is achieved by the experimental proposal of \emph{nanopublication indexes} \cite{kuhn2015iswc}, which are themselves nanopublications. Such indexes can be used to group nanopublications that have trusty URIs using the following command:
\begin{lstlisting}
{c}$ np mkindex trusty.nanopubfile.trig
{o}Index URI: http://np.inn.ac/RAFa_x4h0ng_NXtof35Ie9pQVsAY69Ab3ZQMir2NP8vGc{end}
\end{lstlisting}
The nanopublications of the new index are saved in a file called \texttt{index.trig} unless specified otherwise with the argument \texttt{-o}.

Moving to the part that involves the server network, nanopublications that have trusty URIs (which includes nanopublication indexes) can be published to the network with the following command:
\begin{lstlisting}
{c}$ np publish trusty.nanopubfile.trig
{o}3 nanopubs published at http://np.inn.ac/{end}
\end{lstlisting}
The publication status of a given nanopublication can be checked like this:
\begin{lstlisting}
{c}$ np status -a http://example.org/np1#RAHGB0WzgQijR88g_rIwtPCmzYgyO4wRMT7M91ouhojsQ
{o}URL: http://np.inn.ac/RAHGB0WzgQijR88g_rIwtPCmzYgyO4wRMT7M91ouhojsQ
{o}URL: http://ristretto.med.yale.edu:8080/nanopub-server/RAHGB0WzgQijR88g_rIwtPCmzYgyO{k}...
{o}URL: http://nanopub-server.ops.labs.vu.nl/RAHGB0WzgQijR88g_rIwtPCmzYgyO4wRMT7M91ouhojsQ
{o}URL: http://nanopubs.stanford.edu/nanopub-server/RAHGB0WzgQijR88g_rIwtPCmzYgyO4wRMT7{k}...
{o}URL: http://nanopubs.semanticscience.org/RAHGB0WzgQijR88g_rIwtPCmzYgyO4wRMT7M91ouhojsQ
{o}Found on 5 nanopub servers.{end}
\end{lstlisting}
A given nanopublication that is published on the server network can be retrieved via its URI:
\begin{lstlisting}
{c}$ np get http://www.tkuhn.ch/bel2nanopub/RAhV9IpiUEjbentzGivp1Lbx0BVegp5sgE3BwS0S2RAYM{end}
\end{lstlisting}
All the servers in the network are checked until the nanopublication is found and successfully verified. This command is therefore reliable even if one or several servers are down.
Instead of the complete URI, it is also possible to just specify the trusty URI artifact code:
\begin{lstlisting}
{c}$ np get RAhV9IpiUEjbentzGivp1Lbx0BVegp5sgE3BwS0S2RAYM{end}
\end{lstlisting}
To get the content of a nanopublication index (and not just the top-most index nanopublication), argument \texttt{-c} can be used:
\begin{lstlisting}
{c}$ np get -c -o content.trig RAtF0ivB9B8cb-u3K_zElgmRBxiDwfym1yVBRY6VAyWvE{end}
\end{lstlisting}
Argument \texttt{-o} specifies again the name of the output file.
The remaining commands as introduced above are equally intuitive to use. Just entering the command without any arguments will output a list of all argument options.

\section{Web Interface}

\begin{figure}[tb]
\begin{center}
\includegraphics[width=\textwidth]{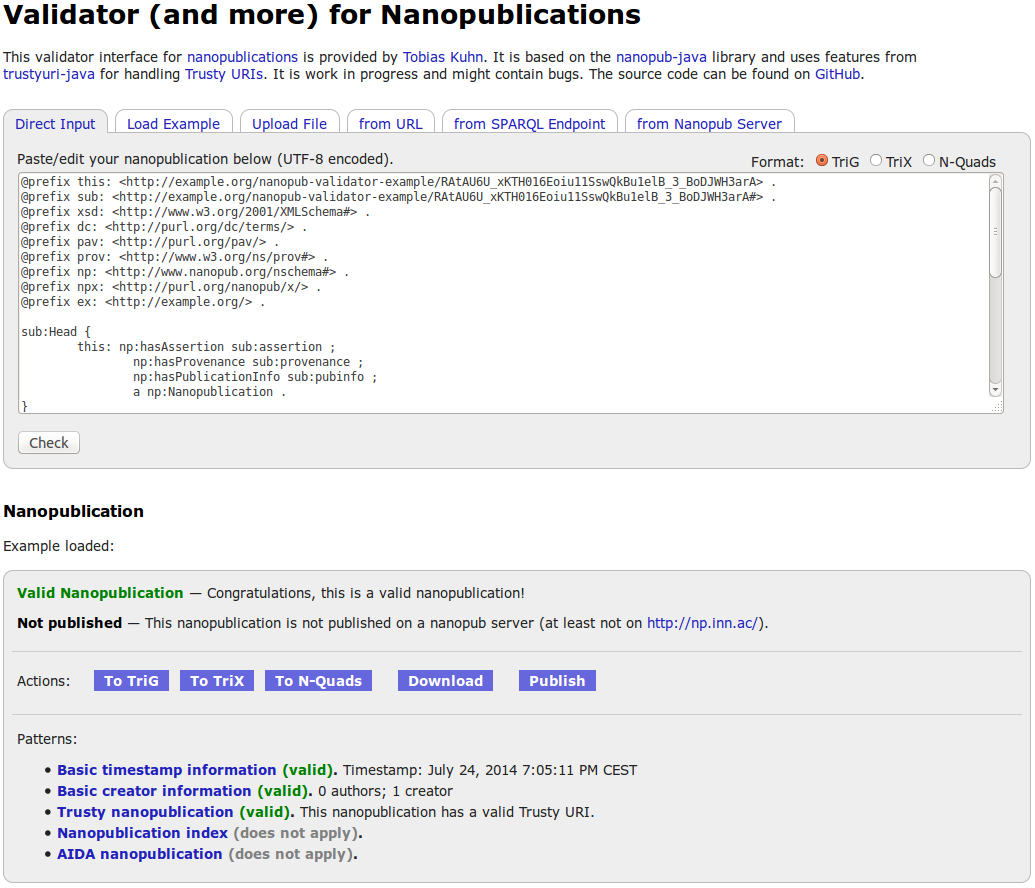}
\caption{This is a screenshot of the nanopublication validator interface at \url{http://nanopub.inn.ac}.}
\label{fig:validator}
\end{center}
\end{figure}

Many of the features described above are made available through the nanopublication validator Web interface,\footnote{\url{https://github.com/tkuhn/nanopub-validator}} an instance of which can be accessed at \url{http://nanopub.inn.ac}. Figure \ref{fig:validator} shows a screenshot. With this interface, the well-formedness of nanopublications can be checked as well as the adherance to a number of patterns. They can furthermore be transformed into different RDF serializations, and published to the server network. Loading of nanopublications is possible via form input, upload, fetching from a URL, SPARQL endpoint access, and retrieval from the server network.
In general, this web interface and the underlying library are supposed to support the development of best practices for the nanopublication community by providing a solid basis for discussion, by allowing for experimental features to be tested and discussed, and by facilitating the implementation of prototypes.

The current server network on which many of the unofficial features depend, can be explored via a monitor interface at \url{http://npmonitor.inn.ac}.

\section{Conclusions}

The \texttt{nanopub-java} library provides a stable implementation of the nanopublication concept, adhering to its specified guidelines. It is openly licensed under the terms of the MIT license, is available on The Central Repository,\footnote{\url{https://search.maven.org/\#artifactdetails|org.nanopub|nanopub|1.7|jar}}
and has so far been used in about a dozen open-source codebases.\footnote{\url{https://github.com/Nanopublication/nanopub-java\#usage-tracking}}

In general, we believe that this library can be a valuable resource for tools that use RDF data in the context of provenance recording, reproducibility, data publishing, data reuse, and reliable retrieval of Linked Data.

\bibliographystyle{abbrv}
\bibliography{nanopubjava}

\end{document}